\documentclass[10pt,a4paper,twocolumn]{article}

\begin{document}
\title{Quantum Mechanical Description of Fluid Dynamics}
\author{\small H. Y. Cui\footnote{E-mail: hycui@public.fhnet.cn.net}\\
\small Department of Applied Physics\\
\small Beijing University of Aeronautics and Astronautics\\
\small Beijing, 100083, China}
\date{\small \today}

\maketitle

\begin{abstract}
\small
In this paper, we deal with fluid motion in terms of quantum mechanics.
Mechanism accounting for the appearance of quantum behavior is discussed.\\ 
\\ \\ \\
\end{abstract}

Consider a ideal fluid which is composed of discrete identical particles,
its mass and charge are $m$ and $q$ respectively, it is convenient to
consider the fluid to be a flow characterized by a 4-vector velocity field $%
u(x_1,x_2,x_3,x_4=ict)$ in a Cartesian coordinate system (in a laboratory
frame of reference). The particle will be affected by the 4-vector force due
to electromagnetic interaction. According to the relativistic Newton's
second law, the motion of the particle satisfies the following governing
equations

\begin{eqnarray}
m\frac{du_\mu }{d\tau } &=&qF_{\mu \nu }u_\nu  \label{1a} \\
u_\mu u_\mu &=&-c^2  \label{1b}
\end{eqnarray}
where $F_{\mu \nu }$ is the curl of electromagnetic 4-vector potential $A$.
Since the Cartesian coordinate system is a frame of reference whose axes are
orthogonal to one another, there is no distinction between covariant and
contravariant components, only subscripts need be used. Here and below,
summation over twice repeated indices is implied in all case, Greek indices
will take on the values 1,2,3,4, and regarding the mass $m$ as a constant.
Eq.(\ref{1a}) and (\ref{1b}) is valid at every point for every particle. As
mentioned above, the 4-vector velocity $u$ can be regarded as a 4-vector velocity
field, then

\begin{equation}
\frac{du_\mu }{d\tau }=\frac{\partial u_\mu }{\partial x_\nu }\frac{dx_\nu }{%
d\tau }=u_\nu \partial _\nu u_\mu  \label{2}
\end{equation}

\begin{equation}
qF_{\mu \nu }u_\nu =qu_\nu (\partial _\mu A_\nu -\partial _\nu A_\mu )
\label{3}
\end{equation}
Substituting them back into Eq.(\ref{1a}), and re-arranging their terms, we
obtain

\begin{eqnarray}
u_\nu \partial _\nu (mu_\mu +qA_\mu ) &=&u_\nu \partial _\mu (qA_\nu ) 
\nonumber \\
&=&u_\nu \partial _\mu (mu_\nu +qA_\nu )-u_\nu \partial _\mu (mu_\nu ) 
\nonumber \\
&=&u_\nu \partial _\mu (mu_\nu +qA_\nu )-\frac 12\partial _\mu (mu_\nu u_\nu
)  \nonumber \\
&=&u_\nu \partial _\mu (mu_\nu +qA_\nu )-\frac 12\partial _\mu (-mc^2) 
\nonumber \\
&=&u_\nu \partial _\mu (mu_\nu +qA_\nu )  \label{4}
\end{eqnarray}
Using the notation

\begin{equation}
K_{\mu \nu }=\partial _\mu (mu_\nu +qA_\nu )-\partial _\nu (mu_\mu +qA_\mu )
\label{5}
\end{equation}
Eq.(\ref{4}) is given by

\begin{equation}
u_\nu K_{\mu \nu }=0  \label{6}
\end{equation}
Because $K_{\mu \nu }$ contains the variables $\partial _\mu u_\nu $, $%
\partial _\mu A_\nu $, $\partial _\nu u_\mu $ and $\partial _\nu A_\mu $
which are independent from $u_\nu $, then a solution satisfying Eq.(\ref{6})
is of

\begin{eqnarray}
K_{\mu \nu } &=&0  \label{7a} \\
\partial _\mu (mu_\nu +qA_\nu ) &=&\partial _\nu (mu_\mu +qA_\mu )
\label{7b}
\end{eqnarray}
The above equation allows us to introduce a potential function $\Phi $ in
mathematics, further set $\Phi =-i\hbar \ln \psi $, we obtain a very
important equation

\begin{equation}
(mu_\mu +qA_\mu )\psi =-i\hbar \partial _\mu \psi  \label{8}
\end{equation}
where $\psi $ representing the wave nature may be a
complex mathematical function, its physical meanings will be determined from
experiments after the introduction of the Planck's constant $\hbar $.

Multiplying the two sides of the following familiar equation by $\psi $

\begin{equation}
-m^2c^2=m^2u_\mu u_\mu  \label{9}
\end{equation}
which is valid at every point in the 4-vector velocity field, and using Eq.(\ref{8}%
), we obtain

\begin{eqnarray}
-m^2c^2\psi &=&mu_\mu (-i\hbar \partial _\mu -qA_\mu )\psi  \nonumber \\
&=&(-i\hbar \partial _\mu -qA_\mu )(mu_\mu \psi )-[-i\hbar \psi \partial
_\mu (mu_\mu )]  \nonumber \\
&=&(-i\hbar \partial _\mu -qA_\mu )(-i\hbar \partial _\mu -qA_\mu )\psi 
\nonumber \\
&&-[-i\hbar \psi \partial _\mu (mu_\mu )]  \label{10}
\end{eqnarray}
According to the continuity condition for the fluid

\begin{equation}
\partial _\mu (mu_\mu )=0  \label{11}
\end{equation}
we have

\begin{equation}
-m^2c^2\psi =(-i\hbar \partial _\mu -qA_\mu )(-i\hbar \partial _\mu -qA_\mu
)\psi  \label{12}
\end{equation}
Its form is known as the Klein-Gordon equation.

On the condition of non-relativity, the Schrodinger equation can be
derived from the Klein-Gordon equation \cite{Schiff}(P.469).

However, we must admit that we are careless when we use the continuity
condition Eq.(\ref{11}), because, from Eq.(\ref{8}) we obtain

\begin{equation}
\partial _\mu (mu_\mu )=\partial _\mu (-i\hbar \partial _\mu \ln \psi
-qA_\mu )=-i\hbar \partial _\mu \partial _\mu \ln \psi  \label{13}
\end{equation}
where we have used the Lorentz gauge condition. Thus from Eq.(\ref{9}) to
Eq.(\ref{10}) we obtain

\begin{equation}
-m^2c^4\psi =(-i\hbar \partial _\mu -qA_\mu )(-i\hbar \partial _\mu -qA_\mu
)\psi +\hbar ^2\psi \partial _\mu \partial _\mu \ln \psi   \label{14}
\end{equation}
This is of a complete wave equation for describing accurately the motion of
the fluid. The Klein-Gordon equation is a linear equation so that the principle 
of superposition remains valid, with the addition of the last term of Eq.(\ref{14}), 
Eq.(\ref{14}) turns to display chaos.

In the following we shall show the Dirac-form equation from Eq.(\ref{8}) and Eq.(\ref{9}). From Eq.(\ref{8}), the wave function can be given in integral form by

\begin{equation}
\Phi =-i\hbar \ln \psi =\int\nolimits_{x_0}^x(mu_\mu +qA_\mu )dx_\mu +\theta
\label{15}
\end{equation}
where $\theta $ is an integral constant, $x_0$ and $x$ are the initial and
final points of the integral with an arbitrary integral path. Since the
Maxwell's equations are gauge invariant, Eq.(\ref{8}) should preserve
invariant form under a gauge transformation specified by

\begin{equation}
A_\mu ^{\prime }=A_\mu +\partial _\mu \chi ,\quad \psi ^{^{\prime
}}\leftarrow \psi  \label{16}
\end{equation}
where $\chi $ is an arbitrary function. Then Eq.(\ref{15}) under the gauge
transformation is given by

\begin{eqnarray}
\psi ^{^{\prime }} &=&\exp \left\{ \frac i\hbar \int\nolimits_{x_0}^x(mu_\mu
+qA_\mu )dx_\mu +\frac i\hbar \theta \right\} \exp \left\{ \frac i\hbar
q\chi \right\}  \nonumber \\
&=&\psi \exp \left\{ \frac i\hbar q\chi \right\}  \label{17}
\end{eqnarray}
The situation in which a wave function can be changed in a certain way
without leading to any observable effects is precisely what is entailed by a
symmetry or invariant principle in quantum mechanics. Here we emphasize that
the invariance of velocity field is hold for the gauge transformation.

Suppose there is a family of wave functions $\psi ^{(j)},j=1,2,3,...,N,$
which correspond to the same velocity field denoted by $P_\mu =mu_\mu $,
they are distinguishable from their different phase angles $\theta $ as in
Eq.(\ref{15}). Then Eq.(\ref{9}) can be given by

\begin{equation}
0=P_\mu P_\mu \psi ^{(j)}\psi ^{(j)}+m^2c^2\psi ^{(j)}\psi ^{(j)}  \label{18}
\end{equation}
Suppose there are matrices $a_\mu $ which satisfy

\begin{equation}
a_{\nu lj}a_{\mu jk}+a_{\mu lj}a_{\nu jk}=2\delta _{\mu \nu }\delta _{lk}
\label{19}
\end{equation}
then Eq.(\ref{18}) can be rewritten as

\begin{eqnarray}
0 &=&a_{\mu kj}a_{\mu jk}P_\mu \psi ^{(k)}P_\mu \psi ^{(k)}  \nonumber \\
&&+(a_{\nu lj}a_{\mu jk}+a_{\mu lj}a_{\nu jk})P_\nu \psi ^{(l)}P_\mu \psi
^{(k)}|_{\nu \geq \mu ,when\nu =\mu ,l\neq k}  \nonumber \\
&&+mc\psi ^{(j)}mc\psi ^{(j)}  \nonumber \\
&=&[a_{\nu lj}P_\nu \psi ^{(l)}+i\delta _{lj}mc\psi ^{(l)}][a_{\mu jk}P_\mu
\psi ^{(k)}-i\delta _{jk}mc\psi ^{(k)}]  \nonumber \\
\label{20}
\end{eqnarray}
Where $\delta _{jk}$ is the Kronecker delta function, $j,k,l=1,2,...,N$. For
the above equation there is a special solution given by

\begin{equation}
\lbrack a_{\mu jk}P_\mu -i\delta _{jk}mc]\psi ^{(k)}=0  \label{21}
\end{equation}

There are many solutions for $a_\mu $ which satisfy Eq.(\ref{19}), we select
a set of $a_\mu $ as

\begin{eqnarray}
N &=&4,\quad a_\mu =\gamma _\mu \quad (\mu =1,2,3,4)  \label{22a} \\
\gamma _n &=&-i\beta \alpha _n\quad (n=1,2,3),\quad \gamma _4=\beta
\label{22b}
\end{eqnarray}
where $\gamma _\mu ,\alpha $ and $\beta $ are the matrices defined in the
Dirac algebra\cite{Harris}(P.557). Substituting them into Eq.(\ref{21}), we
obtain

\begin{equation}
\lbrack ic(-i\hbar \partial _4-qA_4)+c\alpha _n(-i\hbar \partial
_n-qA_n)+\beta mc^2]\psi =0  \label{23}
\end{equation}
where $\psi $ is an one-column matrix about $\psi ^{(k)}$.

Let index $s$ denote velocity field, then $\psi _s(x)$ whose four component
functions correspond to the same velocity field $s$ may be regarded as the
eigenfunction of the velocity field $s$, it may be different from the
eigenfunction of energy. Because the velocity field is an observable in a
physical system, in quantum mechanics we know that the $\psi _s(x)$
constitute a complete basis in which arbitrary function $\phi (x)$ can be
expanded in terms of them

\begin{equation}
\phi (x)=\int C(s)\psi _s(x)ds  \label{24}
\end{equation}
Obviously, $\phi (x)$ satisfies Eq.(\ref{23}). The form of Eq.(\ref{23}) is
well known as the Dirac equation.

Alternatively, another method to show the Dirac equation is more
traditional: At first, we show the Dirac equation of free particle by
employing plane waves, we easily obtain Eq.(\ref{23}) on the condition of $%
A_\mu =0$; Next, adding electromagnetic field, plane waves are valid in any
finite small volume with the momentum of Eq.(\ref{8}) when we regard the
field to be uniform in the volume, so the Dirac equation Eq.(\ref{23}) is
valid in the volume even if $A_\mu \neq 0$, plane waves constitute a
complete basis in the volume; Third, the finite small volume can be chosen
to locate at anywhere, then anywhere have the same complete basis, therefore
the Dirac equation Eq.(\ref{23}) is valid at anywhere.

But, do not forget that the Dirac equation is a special solution of Eq.(\ref
{20}), therefore we believe there are some quantum effects beyond the Dirac
equation.

It follows from Eq.(\ref{8}) that the path of particle is analogous to
''lines of electric force'' in 4-dimensional space-time. In the case that
the Klein-Gordon equation is valid, i.e. Eq.(\ref{11}) is valid, at any
point, the path can have but one direction (i.e. the local 4-vector velocity
direction), hence only one path can pass through each point of the
space-time. In other words, the path never intersects itself when it winds
up itself into a cell about a nucleus. No path originates or terminates in
the space-time. But, in general, the divergence of the 4-vector velocity field does
not equal to zero, as indicated in Eq.(\ref{13}).

In mechanics of fluid, it was shown that vortexes can not originate or disappear in 
an ideal fluid moving in the field of a potential force because of $curl \overrightarrow{v}=0$. 
But according to Eq.(\ref{7b}) and Eq.(\ref{8}), the condition of $curl \overrightarrow{v}=0$ 
is hardly satisfied  by the ideal fluid when the interaction between particles must be 
taken into account, consequently, one can constantly observe the creation 
and disappearance of the vortexes in a real fluid with where the internal or external 
interaction occurs.

Based on the above derivation, we confirm that the dynamic condition of
appearance of quantum behavior in the ideal fluid is that the Planck's
constant $\hbar $ is not relatively small in analogy with that for single
particle, the condition of the appearance of spin structure in the fluid is
that Eq.(\ref{13}) is not negligeable. The mechanism profoundly accounts for
the quantum wave natures\cite{Cui0173}\cite{Cui01114}.

\end{document}